# An Internet of Things Oriented Approach for Water Utility Monitoring and Control

CRISTINA TURCU, CORNEL TURCU, VASILE GAITAN
Computer, Electronics and Automation Department
Stefan cel Mare University of Suceava
13, University Street, 720229, Suceava
ROMANIA
cristina@eed.usv.ro, cturcu@eed.usv.ro, gaitan@eed.usv.ro

*Abstract:* - This paper aims to propose a more efficient distributed monitoring and control approach for water utility in order to reduce the current water loss. This approach will help utilities operators improve water management systems, especially by exploiting the emerging technologies. The Internet of Things could prove to be one of the most important methods for developing more utility-proper systems and for making the consumption of water resources more efficient.

## 1 Introduction

A common characteristic of water demand in urban areas worldwide is its continuing growth over time. Due to different factors like population growth, climate change or even lifestyle changing this demand is expecting to rise in the near future. There are two possible solutions to this problem. The first one is so called "supply-oriented" solution consisting in identifying and exploiting new water resources while the other one is called "demand-oriented" solution that consists in a better manner exploiting of the already available water resources. In the actual condition of water resource diminution the second one become a crucial one. In these circumstances it is the responsibility of every actor playing a role in the water management process to achieve efficient water use.

Functional integration and geographical distribution are two important characteristics associated with recent trends in water supply management. One of the important problems in this area is a technical one, given by the existence of different equipment, devices and software from different suppliers. All these elements must be connected in a proper way so that the entire system can be operable. Another problem is related to the system functionality. A supply water system includes different levels (such as field equipment, process control, management and applications) and its proper design can be seen in the way it assures interoperability of control functions on all its component levels.

The quantitative evaluation of utilities consumption is carried out by counters characteristic for each type of water supplier. In order to monitor the water consumption data, it is necessary that these counters provide output data specific to the metering process. The providers and consumers of utilities have the following characteristics:

- They are geographically distributed, in cities (metropolitan areas) or in larger areas (wide-spread area) (Fig. 1);
- The distribution points are organized hierarchically and can have different complexity level;
- Data acquisition does not require hard real-time facilities, because there is not a closed-loop feedback, the information being used mainly for supervision at a distance, controlling, service, and invoicing. In this situation soft real time conditions may be imposed;
- The access to a medium of communication can be continuous, intermittent, or may not exist at all.

Currently, a high number of systems for water utility monitoring and control are non-interoperable. Many proprietary or semi-closed solutions are implemented, some of them addressing specific issues.

In fact, the control systems used in water distribution utilities are similar to the manufacturing systems, still they have unique aspects. So, we can take into consideration a data acquisition model for water utilities which is derived from the hierarchical levels in enterprises. In Fig. 2 [9] we present the hierarchy of large control systems in enterprises. In the case of water utility monitoring, levels 1 and 2 are mainly geographically distributed, and for monitoring the water utilities, Supervisory Control and Data Acquisition (SCADA) applications are used. Other applications at an enterprise resource planning (ERP) level will carry out operations of billing and tracking the payments of utilities providers. There can be seen that there is vertical flow at the superior levels of the enterprise and distributed flow for field data acquisition.

Supervisory Control and Data Acquisition (SCADA) systems are used in a wide range of processes such as:

- infrastructure processes (e.g. oil and gas pipelines, water management, waste management, electrical power transmission and distribution, etc.),
- industrial processes (e.g. manufacturing, refining, power generation, etc.),
- facilities processes (e.g. buildings, bus terminals, airports, etc.)

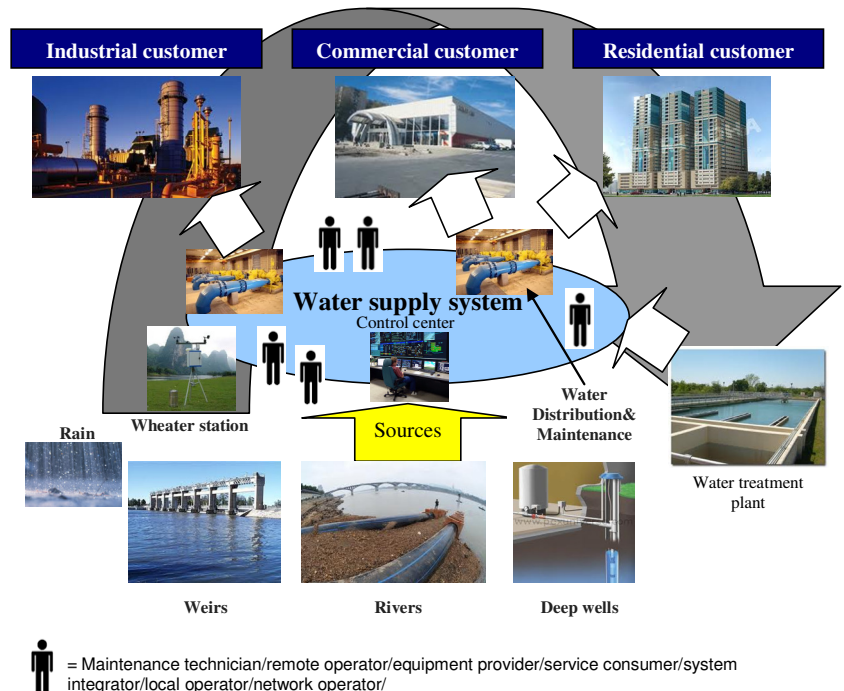

Fig. 1. An example of systems and actors participating in geographically distributed applications

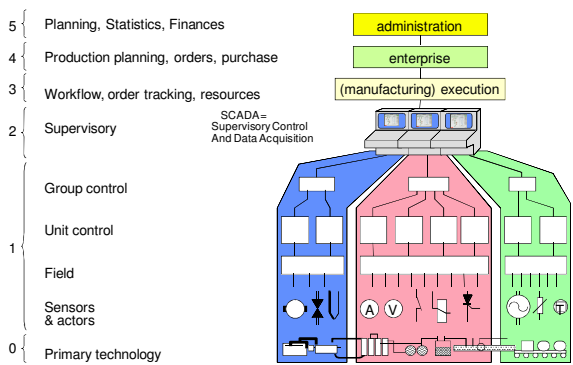

Fig. 2. Large control system hierarchy [9]

The main aim of the SCADA systems is monitoring and controlling different equipments and

devices. As far as water utilities are concerned, some important aspects must be taken into consideration:

- the existence of a wide range of SCADA systems; these systems could be classified depending on the following: the year when they came into operation, their capabilities and accessibility, etc;
- the complexity of the systems rendered by the hardware architecture and programming tools used in their development; this aspect is fully dependent on the deployment group stated by contractors, vendors and even consultants.

The aforementioned aspects have an important impact on the complexity of these kinds of systems. As a result, these systems are often difficult to manage in terms of operations, maintenance, connectivity, and overall robustness [1].

In the last years, new sensors and devices with Internet connection and able to provide real-time information and access have emerged. The technical literature contains many studies describing the benefits of a sensor-based distributed computing infrastructure. However, these studies do not provide an Internet of Thins oriented solution for the development and management of water utilities.

The current paper considers various enabling technologies that could be exploited in order to design and implement an improved water monitoring and control system. Internet of Things is viewed as an evolutionary process, rather than a completely new one. Thus, "from anytime, anyplace connectivity for anyone, we will now have connectivity for anything" [8], not only for anyone. Radio frequency identification (RFID) technology is considered to uniquely identify things (devices, sensors, etc. specific for water utility monitoring and controlling), that should be connected to the Internet of Things. Then, we present a background of multi-agent technologies, that could be used to support the information exchange between various entities of interest. Also, the requirements for an Internet of Things oriented solution for water monitoring and control are presented. Finally, in the last section conclusions are drawn.

## 2 Enabling Technologies
### 2.1 Internet of Things
The integration of sensors, devices, etc. in the Internet is known as Internet of Things (IoT). In the Internet of Things concept, the term "thing" can refer either to people, objects (e.g., devices, sensor, machine, etc.) or information. At the present moment, there are various definitions of the "Internet of Things" varying depending on the context, the effects, and the views of the one giving the definition. Thus, from a things-oriented and an Internet-oriented perspective, the Internet of Things is viewed as "a world where things can automatically communicate to computers and each other providing services to the benefit of the human kind" [4]. According to [5], who regards the Internet of Things from a semantic-oriented perspective, IoT is "a world-wide network of interconnected objects uniquely addressable, based on standard communication protocols". However, there are many common points for most of the definitions of the Internet of Things, such as [6]:

- the ubiquitous nature of connectivity,

- the global identification of every thing,
- the ability of each thing to send and receive data across the Internet or across the private network they are connected into.

According to the identified research agenda for the Internet of Things [7], further research is needed in the development, convergence, and interoperability of technologies for identification and authentication. These technologies can operate at a global scale. Also, there is a need for an open architecture to maximise the interoperability among heterogeneous systems and distributed resources, including providers and consumers of information and services, whether they are human beings, software, smart objects or devices [7].

Radio frequency identification (RFID) technology is considered for this IoT-based approach for water utility monitoring and controlling, proving a low cost solution for the unique identification of things that should be connected to the Internet of Things. This technology will be presented briefly in the following lines.

### 2.2 RFID Technology
Radio Frequency Identification (RFID) technology is an Automatic Identification and Data Capture (AIDC) wireless technology that allows the precise and automatic identification and localization of individual entities (objects, people and animals).

The basic RFID system architecture has two components: contactless electronic tags and an RFID reader. The RFID tag is used to store unique identification data and other specific information related to the tagged entity, whereas the RFID reader allows the reading and writing of these tags. Tags fall into three categories: active (battery-powered), passive (the reader signal is used for activation), or semi-passive (battery-assisted, activated by a signal from the reader). An RFID tag is attached to or embedded in the entity that is to be identified, thus enabling identification, tracking, locating, etc. Moreover, by combining RFID with sensor technology, the number of applications increases tremendously. For example, contactless RFID technology could be used to monitor various physical aspects during the function of an important device.

RFID systems require software, network and database components that enable the information flow from tags to the information infrastructure of an organization, where the information is processed and stored. The systems are application-specific [2].

This paper proposes the use of RFID technology for the identification of entities connected to the Internet of Things; the use of this technology is justified by some of the main characteristics and functionalities of RFID systems:

- Achieving low costs and power efficiency;
- Non-contact and non line-of-sight functionalities enabling data access in harsh environments and through various substances;
- Allowing data storage on RFID tags;
- Allowing the integration of RFID with sensor technology.

In fact, the RFID technology is viewed as a key enabler for the development of the Internet of Things concept.

In order to ensure the exchange of information between different things, we propose to use multi-agent technologies.

## 2.3    Multi-agent Technology

An agent is a software component that has a well-defined role in the operation of a system. Also, an agent must have the ability to communicate with other agents or human users. A multi-agent system is a collection of such entities that cooperate with each other. The multi-agent systems include independent components that communicate in a reactive way; some of them can be instantiated and removed dynamically on demand. By using the multi-agent technology in the implementation of a system, the following advantages could be obtained [3]:

- *High performance*: agents can run in parallel. Thus they can be cloned in the case of very important tasks and goals;
- *High flexibility*: an agent can be developed for any context, providing the interface for different ontologies;
- *High modularity*: the number of connected sources can increase practically without limit.

Thus, we propose to use agent technology in order to solve various problems related to the exchange of information between different system components and different systems implied in water utility monitoring and control.

The monitoring of the water supply consumption can be used to teach citizens how to reduce costs and resources. Future developments for the sewer system can also be taken into consideration with an Internet of Things oriented solution. For example, the Romanian sewer system is a combined one, that is, the same pipes are used for both sanitary wastewater and storm water. Using this type of sewer system causes various problems when it rains, because the system cannot handle the excess flow.

In order to reduce wastewater, an IoT oriented solution can be used to alert people when the sewer is nearing its maximum capacity or it is already overflowing. Also, during drought periods, this solution can be used to inform citizens about the special conditions for saving resources in order to reduce water scarcity.

## 3    The Internet of Things Oriented Approach

We summarize some of the requirements for the next-generation SCADA-based applications that can also be considered for water monitoring and control:

- Real-time;
- Scalability;
- Connectivity – to allow sensors connectivity to enterprise IT systems;
- Support for dynamic environments;
- Security.

An Internet of Things oriented solution must take into consideration all these aspects. Furthermore, it must ensure the autonomy of a variety of IoT entities and resources, such as sensors, smart devices, sensor networks (considering ad-hoc or self-organizing networks), etc.

Water utility monitoring and control systems use the following types of devices:

- *dispersed* devices - that are spread over a wide area;
- *concentrated* devices – that are close to each other.

These devices can be fixed or mobile. The fixed devices (being in fixed locations) might have wired or wireless Internet connection. The mobile devices can be wirelessly connected to the Internet (e.g., by mobile phone).

Currently, there are numerous devices and sensors used in water utility monitoring and control, that are not connected to the Internet. These devices could be connected as things to the Internet of Things in a passive mode through the concentrated devices that are connected to the Internet. In the passive mode, a thing is not connected to the Internet, but can be uniquely identified through an RFID tag. Other Internet-connected things with RFID reading capabilities can identify this thing and publish information related to it on IoT (e.g., thing localization information, sensor data, etc).

In the last years, various solutions have been developed, such as, serial-to-Ethernet converters; these solutions enable the connection of these devices to the Internet, and implicitly the connection in an active mode to the IoT. In the active mode, the thing is connected to the Internet, allowing it to send real-time information to the IoT.

An IoT-based solution must take into account the existing systems and tools for monitoring and controlling the water utility. Despite various limitations imposed by different aspects such as functional capabilities, geographical locations and administrative ownership, it also could propose to extend the current implementations towards internet-connected things. Furthermore, an IoT-based solution must be quickly deployed and easy-to-use, adaptable to a variety of problems common for water utility monitoring and control.

The semantic interactions and the interoperability between various entities in water monitoring and control could be achieved by using a multi-agent system. Thus, the adoption of a multi-agent system for an IoT oriented solution is enabled by a natural mapping between a real world entity and an agent. Regarding the architecture of an IoT oriented solution, the benefits of an approach based on multi-agent systems (instead of using centralized approaches) are listed below:

- The multi-agent system can work even if one agent (as a component of the system) fails or is compromised;
- It is easy to recover a failed agent or replace it with another one;
- Autonomous agents are less vulnerable to attacks than in a centralized architecture.

In fact, an IoT solution can be perceived as a natural extension of the current implementations which must include important additional IoT-based resources and capabilities.

Taking into consideration the above mentioned technologies we propose an IoT-based architecture for water utility monitoring and control (Fig. 3). This architecture is divided into several levels: sensor level, communication level, management and application level, terminal level, and user level.

The sensor level includes different field devices, some of them endowed with computing and communication capabilities. Some devices could act as actuators and based on local sensed conditions or a remote command could initiate proper actions. If these devices are concentrated ones different standards such as CAN, ProfiNet, Modbus etc. could be used to interconnect them.

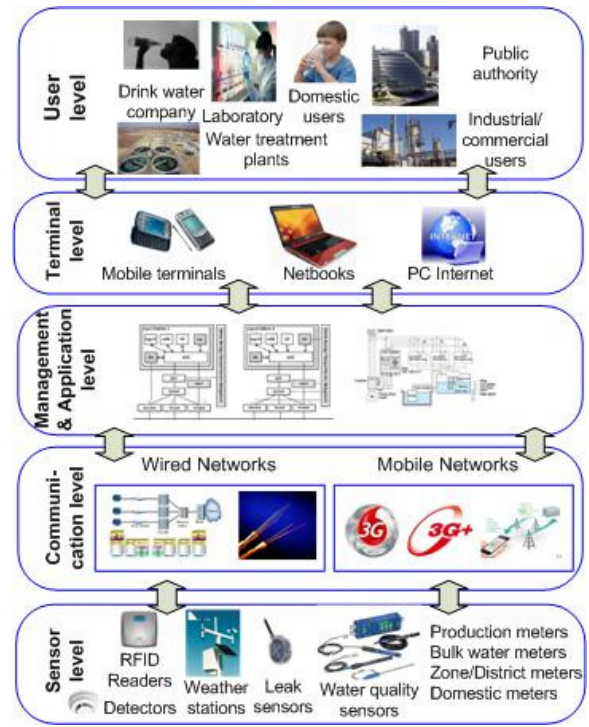

Fig. 3. IoT architecture

The communication level is very important for IoT. This level includes wired (xDSL, optical fibers) and mobile networks (3G, 3G+, GPRS). At this level security and privacy aspects must be treated.

The next level includes intelligent software instruments used to manage, control and operate on IoT devices. It can comprise frameworks, middleware, water monitoring and control applications, etc. Some applications must be designed for modeling processes in water supply. For example for consumption estimation it is possible to develop a model used to provide releases to meet water distribution operation goals. Furthermore, multi-agent technology can bring important benefits when used in designing and deploying applications at this level.

Terminal level consists in different devices used by users from user level.

The Internet of Things becomes more and more interesting in the context of water utility monitoring and control. The IoT approach can provide intelligence to the water and sewer utility, improving economic efficiency and communication with customers. Thus, for example, according to [10], "IoT, because of its ubiquitous sensors and connected systems, will provide authorities with more information and control in order to identify and fix" issues related to leaks and theft.

## 4 Conclusion

Currently, the water monitoring and control is confronted with some issues. Thus, to give some examples, the control systems used by water distribution utilities must operate over a wide geographic area. Worldwide, large water utilities suffer transit looses due to leaks and burst pipes.

The Internet of Things could prove to be one of the most important approaches for developing more utility-proper systems and for making the consumption of water resources more efficient. An IoT solution for water monitoring and control aims at being able to gather data from multiple devices (viewed as things in Internet of Things), analyzing these data and dispatching them or results from processing to various applications or to other devices (also connected to the Internet of Tings).

## Acknowledgments

This paper was supported by the project "Progress and development through post-doctoral research and innovation in engineering and applied sciences – PRiDE – Contract no. POSDRU/89/1.5/S/57083", project co-funded from European Social Fund through Sectorial Operational Program Human Resources 2007-2013.

*References:*

[1] Graham Nasby, Matthew Phillips, SCADA Standardization, Modernization of a Municipal Waterworks with SCADA Standardization: Past, Present, and Planning for the Future, The 6th Annual ISA Water & Wastewater and Automatic Controls Symposium, Missouri, USA, June 22-23, 2011.

[2] OECD, RFID Radio Frequency Identification, OECD Policy Guidance, OECD Ministerial Meeting on the Future of the Internet Economy, Seoul, Korea, 17-18 June 2008.

[3] A. Bouzeghoub and A. Elbyed, A, "Ontology Mapping for Learning Objects Repositories Interoperability", in Intelligent Tutoring Systems, 2006. pp.794-797.

[4] I. Smith, "Internet of Things Around the World", RFID I Congress, Denmark, 2011, Available at: http://www.rfididanmark.dk/fileadmin/Arkiv/Dokumenter/Praesentationer/RFID_i_Danmark_3_maj_2011_-_Internet_of_Things_around_the_world.pdf. Accessed 2012 May 03.

[5] *, "Internet of Things in 2020 – A Road Map for the Future", Available at: ftp://ftp.cordis.europa.eu/pub/fp7/ict/docs/enet/internet-of-things-in-2020-ec-eposs-workshop-report-2008-v3_en.pdf. Accessed 2011 Jan 05.

[6] Advantech, "The Internet of Things, The Future is Connected – Riding the Wave of IoT Growth", Technical White Paper, 2011, Available at: www.advantech-eautomation.com. Accessed 2012 Apr 28.

[7] *, Internet of Things, Strategic Research Roadmap, 2009.

[8] ITU Internet Reports, The Internet of Things, November 2005.

[9] Hubert Kirrmann, Industrial Automation, ABB Research Center, Baden, Switzerland.

[10] Evans D., The Internet of Things - How the Next Evolution of the Internet Is Changing Everything, Cisco Internet Business Solutions Group, April 2011.